\documentclass{PoS}
\usepackage{amsmath}
\usepackage{graphicx}
\usepackage{epstopdf}
\newcommand\apj{ApJ}
\newcommand\aap{A\&A}
\newcommand\apjl{ApJ}
\newcommand\pasj{PASJ}
\newcommand\prd{Phys. Rev. D}

\usepackage{float}
\usepackage{fixltx2e}

\usepackage{amsmath}	% Advanced maths commands
\usepackage{amssymb}	% Extra maths symbols
\newcommand\arcm{\ensuremath{^\prime}}
\newcommand{\pass}{{\sffamily{Pass 8}}}
\title{Deep observations of Cas A with MAGIC indicate it is no PeVatron}
\ShortTitle{Deep observations of Cas A with MAGIC indicate it is no PeVatron}

\author{\speaker{D. Guberman}$^{1}$, J. Cortina$^{1}$, E. de O\~na Wilhelmi$^{2}$, D. Galindo$^{3}$ and A. Moralejo$^{1}$ for the MAGIC collaboration\thanks{We would like to thank the IAC for the excellent working conditions at the ORM in La Palma. We acknowledge the financial support of the German BMBF, DFG and MPG, the Italian INFN and INAF, the Swiss National Fund SNF, the European ERDF, the Spanish MINECO, the Japanese JSPS and MEXT, the Croatian CSF, and the Polish MNiSzW.}\\
       $^{1}$Institut de Fisica d'Altes Energies (IFAE), The Barcelona Institute of Science and Technology, Campus UAB, 08193 Bellaterra (Barcelona), Spain\\
       $^{2}$ Institute for Space Sciences (CSIC/IEEC), E-08193 Barcelona, Spain \\
       $^{3}$ Universitat de Barcelona, ICC, IEEC-UB, E-08028 Barcelona, Spain \\
        E-mail: \email{dguberman@ifae.es}} 

\abstract{It is widely believed that Galactic Cosmic Rays (CR) are accelerated in Supernova Remnants (SNRs) through the process of diffusive shock acceleration. In this scenario, particles should be accelerated up to energies around 1 PeV (the so-called \textit{knee}) and emit gamma rays. To test this hypothesis precise measurements of the gamma-ray spectra of young SNRs at TeV energies are needed. Among the already known SNRs, Cassiopea A (Cas A) appears as one of the best candidates for such studies, because it is relatively young (about 300 years) and it has been largely studied in radio and X-ray bands, which constrains essential parameters for testing emission models. Here we present the results of a multi-year campaign of Cas A observations with the MAGIC Imaging Atmospheric Cherenkov Telescopes between December 2014 and October 2016, for a total of 158 hours of good-quality data. We obtained a spectrum of the source from $\sim$100~GeV to $\sim$10~TeV and fit it assuming it follows a power-law distribution, with and without exponential cut-off. We found, for the first time in the Very High Energy regime (VHE, $\gtrsim$50~GeV), observational evidence for the present of a cut-off in Cas A: an exponential cut-off at about 3.5 TeV is preferred with 4.6 $\sigma$ significance over a pure power-law scenario. Assuming that TeV gamma rays are produced by hadronic processes and that there is no significant cosmic ray diffusion, this indicates that Cas A is not a PeVatron (PeV accelerator) at its present age.}

\FullConference{35th International Cosmic Ray Conference --- ICRC2017\\
		10--20 July, 2017\\
		Bexco, Busan, Korea}

\begin{document}

\section{Introduction}\label{sect:intro}

It is widely accepted that the bulk of the cosmic-rays (CRs), up to energies of 3 PeV (the so called \textit{knee}), are of galactic origin. However, the questions regarding where within the galaxy and how these particles are accelerated remain still unanswered. Supernova remnants (SNRs) have been for a long time the favourite candidates for being a PeVatron, a system capable of accelerating particles up to PeV energies, and for being the main contributors to the galactic CR sea \cite{Berezhko_2003,bell_2013_2,Drury_2014}. This hypothesis is supported by two arguments. On one hand, SNRs can supply the power needed to explain the observed CR energy density if $\sim$10\% of the energy released in a Supernova explosion is transferred to accelerate particles. On the other hand Diffusive Shock Acceleration (DSA, \cite{bell_2013}) offers a plausible acceleration mechanism and explains the CR spectral shape. During the last years, observations at different wavelengths have shown that particles are being accelerated to relativistic energies in SNRs. In particular, observations in the $\gamma$-ray band showed evidence of the presence of relativistic protons in SNRs~\cite{ackermann_2013}. However, no signature of particle acceleration up to PeV energies has yet been found.

Cassiopeia A (Cas~A) is one of the few good candidates to study acceleration of particles to very high energies. It is a relatively young SNR (330 yrs old) and it has been largely studied in radio and X-ray wavelengths. As a result, many parameters of the remnant that are needed to test radiation and acceleration models are reasonably well constrained. It is located at a distance of 3.4$^{+0.3}_{-0.1}$~kpc and has an angular diameter of 5 \arcm{}~\cite{reed_1995}.

Several emission regions have been identified in radio and X-ray bands \cite{Lastochkin_1963,Medd_1965,Allen_1967,Parker_1968,Braude_1969, Hales_1995,Anderson_1991, Gotthelf_2001,suzaku_2009,Grefenstette_2015, integral_2016}. In the gamma-ray domain, $Fermi$-LAT detected the source at GeV energies~\cite{abdo_2010} and later derived a spectrum that displays a
low energy spectral break at 1.72$\pm$1.35 GeV~\cite{yuan_2013}. In the TeV energy range, Cas A was first detected by HEGRA~\cite{aharonian_2001} and later confirmed by MAGIC~\cite{albert_2007}. VERITAS has recently reported a spectrum extending well above 1~TeV~\cite{holder_2016}. The spectrum seems to steepen from the $Fermi$-LAT energy range to the TeV bands. The photon indices measured by HEGRA, MAGIC, and VERITAS are seemingly larger than the Fermi-LAT index of 2.17$\pm$0.09, but the statistical and systematic errors are too large for a firm conclusion.

It has not been established yet which is the nature of the particles that produce the $\gamma$-ray emission observed in Cas A. Recent results from multi-wavelength modelling tend to discard a pure leptonic origin, favouring either an scenario highly dominated by hadronic emission or a more complex scenario involving several mechanisms from more than one population \cite{Berezhko_2003,yuan_2013,Vink_2003,Saha_2014,Zirakashvili_2014}.
This last scenario is supported by the fact that several plausible acceleration regions have been identified in Cas A. 
%Chandra X-ray images \cite{Gotthelf_2001} and high-resolution VLA radio synchrotron maps \cite{Anderson_1995} show a thin outer edge to the SNR that has been interpreted to represent the forward shock where the blast wave encounters the circumstellar medium \cite{DeLaney_2003}. High-resolution observations \cite{Gotthelf_2001,Morse_2004,Patnaude_2007,Helder_2008} also show a reverse shock formed well behind the forward shock that decelerates the impinging ejecta.
Unfortunately, the angular resolution of current Cherenkov telescopes is of the size of the SNR, which makes it hard to identify from which of these regions the main fraction of the observed $\gamma$-ray emission is coming from.

In this work we present an overview of the results reported in \cite{THE_PAPER}, after a multi-year observation campaign on Cas A with the MAGIC telescopes.

%For instance, inverse Compton (IC) contribution, up-scattering the large FIR photon field of Cas A itself (with energy density of $\sim$2 eV/cm$^3$ and temperature of 97 K, \cite{Mezger_1986}), is more significant in a region of lower magnetic field, as otherwise it would be suppressed due to fast cooling of electrons.

\section{Observations and data analysis}
\label{sect:observations}

%\subsection{The MAGIC Telescopes}

MAGIC is a system of two 17~m diameter Imaging Atmospheric Cherenkov Telescopes (IACTs), located at an altitude of 2200~m a.s.l. at the Roque de los Muchachos Observatory on the Canary Island of La Palma, Spain (28$^\circ$N, 18$^\circ$W). The cameras of the telescopes consist of photomultiplier tubes (PMTs) that can detect the Cherenkov light that is emitted due to the interaction of very high energy (VHE, $\gtrsim$50~GeV) $\gamma$-ray photons with the atmosphere \cite{Magic_performanceII}.

Observations were performed between December 2014 and October 2016, for a total observation time of 158 hours after data quality cuts. The data correspond to zenith angles between 28 and 50 degrees and most of them ($\sim$73\%) were taken during moonlight time (see Table \ref{tabTime}), under background-light levels that could be up to 12 times brighter than during dark nights. A significant part of the data ($\sim$24\%) were obtained under Reduced High Voltage (HV) settings: the gain of the PMTs is lowered by a factor $\sim$1.7 to decrease the damage inflicted by background light on the photodetectors during strong moonlight time. A detailed study of the performance of the MAGIC telescopes under moonlight has been performed in \cite{MAGIC_moon}.

The data have been analysed following the optimised moonlight analysis described in \cite{MAGIC_moon}. For the spectrum reconstruction a point-like source was assumed and typical selection cuts with 90\% and 75\% $\gamma$-ray efficiency for the $\gamma$-ray/hadron separation and sky signal region radius, respectively \cite{Magic_performanceII}. Three OFF regions were considered for the background estimation.

 \begin{table}
\centering
 \caption{Effective observation time of the different hardware and sky brightness conditions under wich Cas A samples were taken.}\label{tabTime}
\begin{tabular}{ c  c }
\hline
% & & \\
Observation conditions & Time [h] \\
\hline
   \hline

     Dark and Nominal HV & 42.2 \\
     Moon and Nominal HV & 77.7 \\     
     Moon and Reduced HV & 38.1 \\     
   \hline
   All configurations & 158.0 \\
 \end{tabular}
 \end{table}

%\subsection{$Fermi$-LAT}

We also analysed 8.3~yr of LAT data (up to December 6, 2016) on a $15^\circ \times 15^\circ$ region around the position of Cas A, using the usual filters recommended by the $Fermi$-LAT collaboration. The data set was reduced and analysed using \emph{Fermipy}\footnote{http://fermipy.readthedocs.org/en/latest/}, a set of python tools which automatise the \pass{} analysis.
%(removing intervals when the rocking angle of the LAT was greater than 52$^\circ$ or when parts of the region-of-interest, ROI, were observed at zenith angles larger than 90$^\circ$). In order to derive the energy spectrum we applied a maximum likelihood estimation analysis in 13 independent energy bins from 60\,MeV to 500\,GeV, modelling the Galactic and isotropic diffuse emission using the templates provided by the $Fermi$-LAT collaboration\footnote{\emph{gll\_iem\_v06.fits} and \emph{iso\_P8R2\_ULTRACLEANVETO\_V6\_v06.txt},
%{http://fermi.gsfc.nasa.gov/ssc/data/analysis/documentation/Cicerone}}. During the broad-band fit, all sources in the third $Fermi$ catalog (3FGL) within the ROI were included. A source located $\sim$3.7$^\circ$ away from Cas A at (l,b)=(113.6$^\circ$,1.1$^\circ$) was added during the fitting process to account for a significant residual excess (with TS$=45.08$).
Following the results obtained by \cite{yuan_2013} we used a smoothly broken power-law function to fit the broadband spectrum of Cas~A ($dN/dE = N_{\rm o}(\frac{E}{E_o})^{\Gamma_1}(1+(\frac{E}{E_b})^{(\Gamma_1-\Gamma_2)/\beta})^{-\beta}$)  with the parameter $\beta$ fixed to 1 and the energy break to $E_
{\rm b}$=1.7 GeV. E$_{\rm o}$ is the normalisation energy, fixed to 1~GeV. The SED was obtained by fitting the source normalisation factor in each energy bin independently using a power-law spectrum with a fixed spectral index of 2. For each spectral point we required at least a TS of 4, otherwise upper limits at the 95\% confidence level were computed.

\section{Results}

\begin{figure}[t]
\centering
\includegraphics[width=0.8\columnwidth]{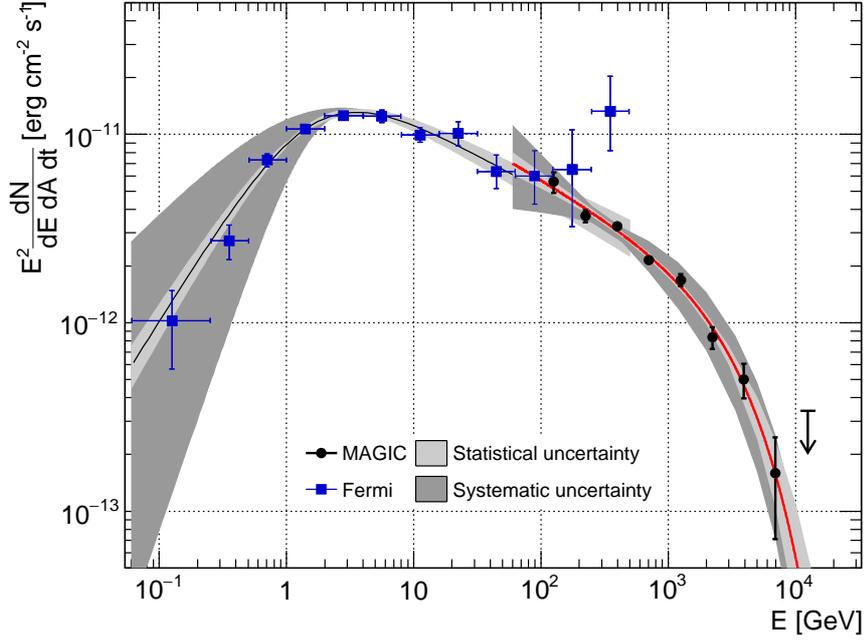}
\caption{Spectral energy distribution measured by the MAGIC telescopes (black dots) and Fermi (blue squares). The red solid line shows the result of fitting the MAGIC spectrum with Eq. \ref{Eq:fit_EPWL}. The black solid line is the broken power-law fit applied to the Fermi spectrum.}\label{fig:SED}
\end{figure}

Figure \ref{fig:SED} shows the gamma-ray spectrum of Cas A obtained with the MAGIC telescopes and $Fermi$. The spectrum obtained by MAGIC can be described with a power-law with an exponential cut-off (EPWL):

\begin{equation}\label{Eq:fit_EPWL}
\dfrac{dN}{dE} = N_0 \left( \dfrac{E}{E_0}\right)^{-\Gamma} \text{exp}\left(- \dfrac{E}{E_c} \right)
\end{equation}

with a normalisation constant $N_0=(1.1 \pm 0.1_{\textit{stat}} \pm 0.2_{\textit{sys}}) \times 10^{-11}$~$\text{TeV}^{-1} \text{cm}^{-2} \text{s}^{-1}$ at a normalisation energy $E_0 = 433~\text{GeV}$, a spectral index $\Gamma = 2.4 \pm 0.1_{\textit{stat}} \pm 0.2_{\textit{sys}}$ and a cut-off energy $E_c = 3.5\left(^{+1.6}_{-1.0}\right)_{\textit{stat}} \left(^{+0.8}_{-0.9}\right)_{\textit{sys}}$~TeV. The spectral parameters of the tested models $\theta=\left\lbrace N_0,\Gamma,E_c \right\rbrace$ are obtained via a maximum likelihood approach.
%The data inputs are the numbers of recorded events (after background suppression cuts) in each bin of estimated energy $E^{i}_{\text{est}}$, both around the source direction ($N^{\text{ON}}_{i}$) and in the three OFF regions ($N^{\text{OFF}}_{i}$). An additional set of nuisance parameters $\mu_{i}$ for modelling the background are also optimized in the likelihood calculation. In each step of the maximisation procedure the expected number of gammas in a given bin of estimated energy ($E_{\text{est}}$) is calculated by folding the gamma spectrum with the MAGIC telescopes response (energy-dependent effective area and energy migration matrix). The background nuisance parameters and the statistical uncertainties in the telescopes response are treated as explained in \cite{Rolke_2005}.
The probability of the EPWL fit is 0.42. We tested the model against the null hypothesis of no cut-off, which is described with a pure power-law (PWL). The probability of the PWL fit is $6 \times 10^{-4}$. A likelihood ratio test between the two tested models favours the one that includes a cut-off at $\sim3.5$~TeV with 4.6$\sigma$ significance.

%Figure \ref{fig:Res} compares the fit residuals for the two tested models: PWL and EPWL. The residuals are here defined as $N^{\text{obs}}_{\text{ON}}/N^{\text{exp}}_{\text{ON}}-1$, where $N^{\text{obs}}_{\text{ON}}$ is the number of observed events (including background) in the ON region and $N^{\text{exp}}_{\text{ON}}$ is the number of events predicted by the fit in the same region. All the bins in estimated energy which contain events are used in the fits, but only those with a \Revision{$2\sigma$} significance gamma-ray excess are shown as SED points in upper panel of Fig. \ref{fig:SED}.

For the $Fermi$-LAT analysis, a broken power-law function with normalisation N$_{\rm o}=(8.0\pm0.4)\times10^{-12}$~ $\text{TeV}^{-1} \text{cm}^{-2} \text{s}^{-1}$, indices ${\Gamma_1=0.90\pm0.08}$ and ${\Gamma_2=2.37\pm0.04}$ is obtained.

%+/- 15%
The systematic uncertainty due to an eventual mismatch on the absolute energy scale between MAGIC data and MC simulations was constrained to be below 15\% in \cite{Magic_performanceII}. By conservatively modifying the absolute calibration of the telescopes by $\pm 15\%$, and re-doing the whole analysis, we can evaluate the effect of this systematic uncertainty in the estimated source spectrum. This does not produce a simple shift of the spectrum along the energy axis, but changes also its hardness. Even in the unlikely scenario in which, through the 158 h of observations, the {\it average} Cherenkov light yield was overestimated by 15\% relative to the MC, by applying the corresponding correction the resulting spectrum is still better fit by an EPWL at the level of 3.1$\sigma$. In the scenario in which the light yield was underestimated in average by 15\%, the EPWL is preferred over the PWL at the 6.5$\sigma$ level.
The systematic uncertainties in the flux normalization and spectral index during moonlight observations were reported in \cite{MAGIC_moon}. 

%The light gray shaded area shows the statistical errors of the obtained broken power-law fit whereas the dark one marks the uncertainty coming from the imperfectness in the Galactic diffuse emission modelling, dominating the Cas A flux uncertainties at low energies. The later were obtained by modifying the galactic diffuse flux by $\pm$6\%. Note that the systematic error due to the diffuse background is greatly reduced above 300 MeV. %Fermi and MAGIC spectra connect smoothly near 100~GeV.

%\begin{figure}
%\includegraphics[width=\columnwidth]{Figures/residuals_EPWLvsPWL}
%\caption{Relative fit residuals for the two tested models fitting MAGIC spectrum: power-law with exponential cut-off (EPWL, upper panel) and power-law (PWL, lower panel). \Revision{The error bars are calculated such that they correspond to the total contribution of each estimated energy bin to the final likelihood of the fit}}\label{fig:Res}
%\end{figure}

\section{Discussion}

With MAGIC observations we have obtained the most precise spectrum of Cas A to date. This study, presented in detail in \cite{THE_PAPER}, reports for the first time observational evidence for the presence of a cut-off in the VHE spectrum of Cas A. The spectrum measured with the MAGIC telescopes can be described with an EPWL with a cut-off at $\sim$3.5 TeV, which is preferred over a PWL scenario with 4.6$\sigma$ significance. 

These high precision measurements allow us to set stronger constrains on the nature of the particles that are responsible for the TeV emission. With this aim, we have used the radiative code and Markov Chain Monte Carlo fitting routines of $naima$\footnote{https://github.com/zblz/naima} \cite{Zabalza_2015}, deriving the present-age particle distribution. The code uses the parametrisation of neutral pion decay by \cite{Kafexhiu_2014}, the parametrization of synchrotron radiation by \cite{NaimaSyn} and the analytical approximations to IC up-scattering of blackbody radiation and non-thermal bremsstrahlung developed by \cite{Khangulyan_2014} and \cite{Baring_1999}, respectively.

We first considered the possibility of a pure leptonic scenario in which the gamma-ray emission is originated by an electron population, described with a power-law with an exponential cut-off function, producing Bremsstrahlung and IC radiation in the gamma-ray range, and synchrotron radiation at lower energies. The photon fields that contribute to the inverse Compton component are the ubiquitous 2.7 K cosmic microwave background (CMB) and the large far infrared (FIR) field measured in Cas A, with a value of $\sim$2 eV/cm3 at 100 keV. 
With the photon field fixed, we searched for the highest possible density of electrons that are compatible with the VHE flux. Once this was set, we could constrain the maximum magnetic field for which the synchrotron radiation produced by the derived population does not exceed the radio and X-ray measurements\footnote{This constraint is due to the fact that, as reported in section~\ref{sect:intro}, several emission regions, likely associated to different particle populations, were identified at those wavelengths.}. 

\begin{figure}[t]
\centering
\includegraphics[width=0.8\columnwidth]{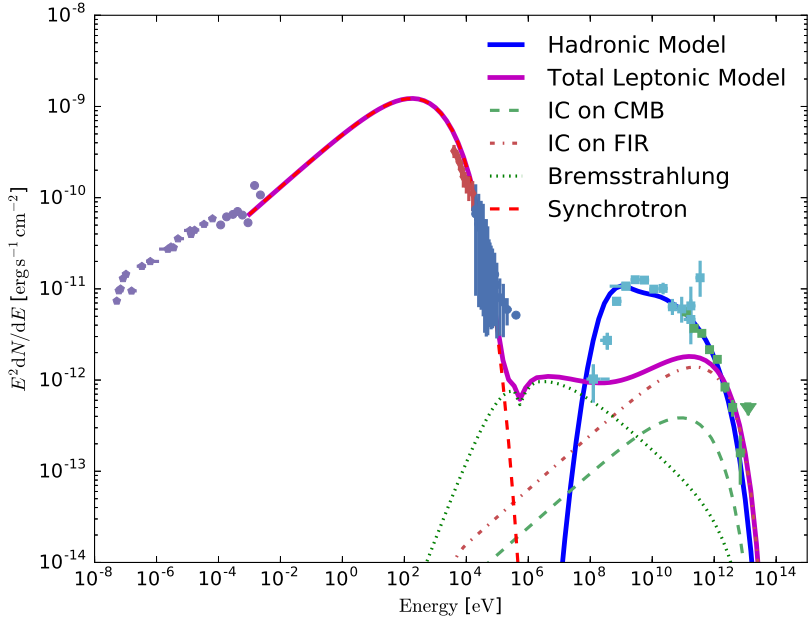}
\caption{Multi-wavelength SED of Cas~A. The different lines show the result of fitting the measured energy fluxes using $naima$ and assuming a leptonic or a hadronic origin of the GeV and TeV emission.}\label{fig:SEDModel}
\end{figure}

The resulting emission from the leptonic model is shown in Fig. \ref{fig:SEDModel}, among with the radio measurements displayed in purple dots \cite{Lastochkin_1963,Medd_1965,Allen_1967,Parker_1968,Braude_1969,Hales_1995,Planck_2014}, soft SUZAKU X-rays marked in red \cite{suzaku_2009} and hard INTEGRAL X-rays in blue \cite{integral_2016}. LAT points are shown in cyan and the MAGIC ones in green. The electron population responsible for the emission that results from the model has an amplitude at 1~TeV of 2$\times$10$^{34}$eV$^{-1}$, spectral index 2.4 and a cut-off energy at 8~TeV. The maximum allowed magnetic field was found to be at B$\simeq$180~$\mu$G, a value that was used to derive the synchrotron emission. The model is generally compatible with the X-ray and radio measurements and it could actually explain the MAGIC spectrum above a few TeV. But it fails to reproduce the $\gamma$-ray spectrum between 1 GeV and 1 TeV, being a factor 2-3 below the measured LAT spectrum. In addition, the derived emission was obtained for a relatively low magnetic field intensity. To accommodate a magnetic field of the order of $\sim$1~mG, as reported in \cite{Uchiyama_2008}, the amplitude of the electron spectrum would need to be decreased at least by a factor 10, rendering a negligible IC contribution at the highest energies. We conclude that a non-negligible contribution from electrons to the TeV emission can only be sustained if it is originated in a region of low magnetic field like the reverse shock.

We also studied the possibility of a pure hadronic scenario. We assumed a population of CRs characterised with a power-law function with an exponential cut-off to fit the gamma-ray data from 60 MeV to 15 TeV, and a target proton density of 10 cm$^{-3}$ \cite{Laming_2003}. The proton spectrum is best-fit with a hard index of 2.21 and an exponential cut-off energy of 12~TeV, which implies a modest acceleration of CRs to VHE, well below the energy needed to explain the CR $knee$. The proton energy above 1 TeV is 5.1$\times$10$^{48}$ erg, which is only $\sim$0.2\% of the estimated explosion kinetic energy of ${\rm E_{\rm sn}} = 2\times10^{51}$ erg \cite{Laming_2003}. The total energy stored in protons above 100~MeV amounts to $9.9\times10^{49}$ erg.

The flat spectral index is in agreement with the standard theory of diffuse shock acceleration, but the low cut-off energy implies that Cas A is an extremely inefficient in the acceleration of CRs at the present moment. Alternatively, Cas A may also be located in a very diffusive region of the Galaxy, resulting in a very fast escape of protons of TeV and higher energies.
This result implies that even if all the TeV emission was of hadronic origin, Cas A could not be a PeVatron at its present age.

%\section{Conclusions}

Several emission regions must be active to explain the radio, X-ray, GeV and TeV emission of Cas A. A purely leptonic model cannot explain the GeV-TeV spectral shape derived using LAT and MAGIC data, as previously suggested based on observations at lower energies 
\cite{Saha_2014,Zirakashvili_2014,Atoyan_2000_1, Atoyan_2000_2}. A leptonic population is undoubtedly necessary to explain the emission at radio and X-ray energies. Indeed the bright steep-spectrum radio knots and the bright radio ring, demand an average magnetic field of $\sim$1 mG \cite{Vink_2003}, whereas the faint plateau surrounding Cas A, seen in Chandra continuum images, is consistent with a lower magnetic field, which might contribute to the observed emission above 1~TeV. 

However, the bulk of the HE and VHE $\gamma$-rays must be of hadronic origin. Cas A is most likely accelerating CRs, although to a rather low energy of a few TeV. Even if some leptonic contribution at VHE produced by IC cannot be excluded, this would not affect our conclusion that acceleration in Cas A falls short of the energies of the $knee$ of the CR spectrum.

%A detailed study of the cut-off shape would provide crucial information to properly understand the reasons of the low acceleration efficiency found. A better understanding of the morphology in which it could be identified which regions of Cas A are the major contributors to the TeV emission would provide new insights on the acceleration processes in the SNR.

%\section*{Acknowledgements}
%-------------------------------------------------------------------

\end{document}